\documentclass{article}
\usepackage{spconf,amsmath,graphicx}

\usepackage{amsmath,amssymb}

\usepackage{pgfpages}
\usepackage{adjustbox}
\usepackage{graphicx}
\usepackage{pdfpages}

\usepackage{array}

\usepackage{dblfloatfix}

\usepackage[compact]{titlesec}
\titlespacing{\section}{0pt}{3ex}{2ex}
\titlespacing{\subsection}{0pt}{2ex}{1ex}
\titlespacing{\subsubsection}{0pt}{2ex}{1ex}

\usepackage{url}

\usepackage{algorithm,algpseudocode}
\usepackage{balance}

\algnewcommand{\Inputs}[1]{%
  \State \textbf{Inputs:}
  \Statex \hspace*{\algorithmicindent}\parbox[t]{.8\linewidth}{\raggedright #1}
}
\algnewcommand{\Initialize}[1]{%
  \State \textbf{Initialize:}
  \Statex \hspace*{\algorithmicindent}\parbox[t]{.8\linewidth}{\raggedright #1}
}

\usepackage{xxcolor}

\usepackage[compact]{titlesec}
\usepackage[left=2cm,top=1cm,right=2cm,nohead,nofoot]{geometry}

\definecolor{bblue}{HTML}{6328FA}
\definecolor{rred}{HTML}{C0504D}
\definecolor{ggreen}{HTML}{9BBB59}
\definecolor{ppurple}{HTML}{9F4C7C}
\definecolor{bbrown}{HTML}{9E600E}

\algdef{SE}[SUBALG]{Indent}{EndIndent}{}{\algorithmicend\ }%
\algtext*{Indent}
\algtext*{EndIndent}
\title{An Analysis of Rhythmic Staccato-Vocalization Based on
  Frequency Demodulation for Laughter Detection in Conversational
  Meetings}
\makeatletter
\def\name#1{\gdef\@name{#1\\}}
\makeatother 
\name{Sucheta Ghosh$^1$, Milos Cernak$^1$, Sarbani Palit$^2$, and B. B. Chaudhuri$^2$}
\address{$^1$Idiap Research
  Institute, Switzerland, $^2$Indian~Statistical~Institute, Kolkata, India.\\
  {\small \tt \{sucheta.ghosh,milos.cernak\}@idiap.ch, \{sarbanip,bbc\}@isical.ac.in}
  }
\begin{document}
%\ninept
%
\maketitle
\begin{abstract}\vspace{0.5cm}
Human laugh is able to convey various kinds of meanings in human
communications. There exists various kinds of human laugh signal, for
example: vocalized laugh and non vocalized laugh. Following the
theories of psychology, among all the vocalized laugh type, rhythmic
staccato-vocalization significantly evokes the positive responses in
the interactions. In this paper we attempt to exploit this observation
to detect human laugh occurrences, i.e., the laughter, in multiparty
conversations from the AMI meeting corpus. First, 
we separate the high energy frames from speech, leaving out the low
energy frames through power spectral density estimation. We borrow the
algorithm of rhythm detection from the area of music analysis to use
that on the high energy frames. Finally, we detect rhythmic laugh
frames, analyzing the candidate rhythmic frames using statistics. This
novel approach for detection of `positive' rhythmic human laughter
performs better than the standard laughter classification baseline.
%mc150915 - moved this above
%% We use AMI meeting corpus for this task.
\end{abstract}
  \section{Introduction}

%    Human laugh plays an important role in human communication due to its inner meanings. Laugh in human speech signal also important in spontaneous automatic speech recognition to increase the word accuracies.   

Human laugh is a crucial social signal due to the range of inner
meanings it carries. This social signaling event \cite{vincia2009} may
denote the topical changes, communication synchrony and positive
affect; on the other hand, it may also show disagreement or satirist
views. Therefore, automatic human laugh occurrence or laughter
detection in speech may have 
many applications in spoken dialog and discourse analysis. In
addition, the detection of this speech event may lead to increase in
the word accuracies in the spontaneous automatic speech recognition.

%\note{I have confusions with ``laugh'' and ``laughter'' -- you use both terms simultaneously. (SG22/06: tried to make up confusion, will revise more times)} 

Human laugh is developed as an inarticulate utterance to serve as an
expressive-communicative social signal. The entire laugh period
generally persists from 2 seconds to 8 seconds \cite{ruch2001}. There
exists many types of laugh. From the acoustical point of view, the
sound of laugh can be voiced, as well as it can be unvoiced, resulting
into the vocalized and non-vocalized laughter. The whole laugh episode
is constituted with a mixture of vocalized and non-vocalized
laugh. 
% The study of Bachorowski et al (2001)
% MC 270815
It was found that the voiced and rhythmic laughs were significantly
and more likely to elicit positive responses than the variants such as
unvoiced grunts and snort like sounds \cite{jab2001}.

The laugh sound or laugh bout can be segmented into three parts,
viz.(1) onset: explosive laugh, short and steep, (2) apex: vocalized
part of laugh and (3) offset: post-vocalized fading part of laugh.
%\begin{enumerate}
%\item onset: explosive laugh, short and steep,
%\item apex: vocalized part of laugh and
%\item offset: post-vocalized fading part of laugh.
%\end{enumerate}
The vocalized apex part is composed of laugh cycles (for example, the
laugh sound ``ha ha''), each cycle is composed of laugh pulses. The
number of pulses depends on the power of the lungs, it can be 4 to 12
for one cycle. These laugh pulses have a rhythmic pattern.

Although it is found that in sustained laughter the apex might be
interrupted by inhalations \cite{ruch2001}, human laughter is easily
recognized through the detection of apex part \cite{drack2009}.
% This work tries to capture this rhythm of the vocalized laugh
% pulses.
Therefore, it is clear that the sound of laugh may also be based on
the rhythmic breathing resulting in a staccato vocalization, i.e., the
vocalization with each sound or note sharply detached or separated
from the others. Detection of the apex part plays dominant role to
recognize the human laughter.

The majority of the previous works on laugh detection
(cf. Section~\ref{sec:relwrk}) follow the supervised classification
paradigm that may face a long extent of a training phase with
considerable amount of costly annotated data. We hypothesize that a
rhythmic nature of the vocalized laugh can allow us to use existing
rhythm-detection signal processing based techniques, e.g., detection
the rhythm in music \cite{sethares2007}, also for the laugh detection.
This would lead to an unsupervised and less data-dependent laugh
detection, as an alternative to conventional machine learning
approaches.

In this work, we propose a three stage procedural method to detect
human laugh using rhythm, through the laugh apex, which is the most
prominent laugh part. This procedural method works in three basic
procedural sequences: first we filter out low Power Spectral Density
(PSD) frames using an automatic PSD threshold computation based on
well-established Otsu's threshold technique \cite{otsu1975}. Then we
analyze all high-energy PSD frames to detect the rhythmic frames (such
as rhythmic speech and/or rhythmic laugh) with a music rhythm detector
algorithm \cite{sethares2007}
%mc150915 - maybe one more sentence here explaining our motivation to
%use frequency demodulation
based on frequency demodulation.
%mc150915
We select higher energy frames because
human laugh is predominantly conceptualized as vowel-like high energy
bursts \cite{jab2001}. Finally, we compute a statistical threshold to
detect only the rhythmic laugh frames.
%mc150915 
We demonstrate the proposed detection method on naturally occurring
conversations, that usually contain plenty of instances of happy and
natural human laugh. Therefore, we choose multiparty meeting
conversations AMI \cite{ami2005} as a database for evaluation. The
recordings of the AMI meeting corpus show a huge variety of
spontaneous expressions.
%mc150915
%% We choose multiparty meeting
%% conversations AMI \cite{ami2005} as a database to demonstrate this
%% approach. The recordings of the AMI meeting corpus show a huge variety
%% of spontaneous expressions, and since it consists of naturally
%% occurring conversations, we find there plenty of instances of happy
%% and natural human laugh.
% First, we detect the human speech frames with high energy. Then, we
% check that whether the high energy frames are rhythmic or not. At
% last, the rhythmic high energy frames are statistically analyzed to
% detect the frames with human laugh occurrences.
% \subsection{Our Contribution}
% \textit{Hereby we do not
%  attempt to detect all kinds of laughter, we target to detect only
%  positive vocalized laughter}.
% The algorithms of the
% first two procedures are already published work for different tasks;
% we use those in this novel task to accomplish our goal.

%\note{Why we filter out the low energy frames? We should introduce
  %briefly with one sentence self computed PSD. --addressed on 26.08.15}

The organisation of this paper is structured as follows: in the next
Section~\ref{sec:relwrk}, we clarify the laugh as a signal and its
types as established by the studies, and also describe in details the
related works on laughter detection and recognition. In the following 
Section~\ref{sec:method} we illustrate the proposed method. In the
next consequent Section~\ref{sec:detexp} we describe used data and
experimental set-ups; this is followed by the discussion about the
results in the subsection~\ref{sec:res}. Finally, we conclude the findings
and possible future works in the Section~\ref{sec:concl}.

\section{Background}
\label{sec:relwrk}

\subsection{The laugh apex}

% Human laugh is a powerful audio-visual expressive-communicative social
% signal. This is basically an inarticulate speech sound. There exists
% vocalized laughter and non-vocalized laughter. At times 

The vocalized laugh can be spontaneous or voluntary. With clinical
observation there is a clear distinction between spontaneous and
voluntary laugh \cite{bright1986}. It is seen that during
spontaneous laugh human self-awareness and self-attention is
diminished. On the other hand, in voluntary laugh human produce a
laugh sound pattern similar to the spontaneous laugh but still it
differs in many aspects like vowel used (viz. the derivative of
schwa), pitch, frequencies and amplitudes, voice quality etc. All
these differences have effects on the rhythm of spontaneous and
voluntary laugh.

Bachorowski et al (2001) found that the vocalized laugh is rhythmic
compared to the snort-like laugh or grunts; and also the vocalized laugh
elicits positive emotion than the other kinds of laugh
\cite{jab2001}. Devillers et al (2007) found that the unvoiced laughs express more negative emotion, whereas the voiced laugh segments are perceived as positive ones \cite{devillers2007}. 

% \note{We need here more review related to the positive/negative vocalised laugh. How they are different?--addressed adding one sentence on 260815}

\subsection{Laughter Detection}
\label{sec:relwrk-detection}

A sizable number of previous works in laugh occurrence detection are
already proved to be impressive in terms of techniques, results and
large set of intricate features \cite{schuller2013}.  The majority of
the previous works follow the supervised classification paradigm that
may face a long extent of training phase with considerable amount of
costly annotated data. Many of these works consider the task as a
binary (i.e. laugh vs. non-laugh) classification
\cite{kennedy2004,truong2007,salamin2013} or segmentation problem
\cite{knox2007, petridis2011}.

The label-type of laughter in the laughter detection tasks may vary
from the coarse-grained label to the fine-grained one. The
coarse-grained laugh detection generally implies to the binary
classification (i.e. laugh vs. non-laugh). There exists some
instances of coarse-level multi-class detection viz. the laughter
classification along with the other non-laugh classes like
silences, fillers etc. \cite{salamin2013}. In other works, it detects
many kinds of laugh such as polite, mirthful, derisive
vs. non-laugh \cite{tanaka2014}.

% \note{??The laughter-type hierarchy for each
% corpus is unique for each corpus; that is a difficulty for us to align
% with several databases.-- Should we remove such a strong comment? on 26.08.15}
    
There also exists a few works on unsupervised classification of the
laugh. Some unsupervised techniques depend on the burst detection and
classification of the burst as laughter \cite{peleg2013}. The affect
bursts are defined as short, emotional and non-speech expressions that
interrupt speech such as respiration, laughter or unintelligible vocal
sound \cite{ms2003}. Therefore it is hard to tag the right meaning of
(single or n-tuple) affective bursts without any reference. The
non-parametric statistical methods also have been exploited in
real-time, training-free framework to detect laughter; still one needs
to extract features for this technique \cite{chou2012}. Majority of
unsupervised methods are primarily tested on their own collected
data.

% \note{What is link between already defined parts of laugh, onset,
%   apex, offset, and the affect bursts? Why we now start talking about
%   emotions? Section II.A needs to introduce these links to emotions.==comment on 260815 Need to discuss}
    
In this work we attempt to propose a real-time, rhythm-based approach
for laughter detection. We attempt to exploit the rhythmic pattern of
laughter, following the work of Bachorowski et al (2001)
\cite{jab2001}, we aim to detect the vocalized laughter through
detection of the laugh apex occurrence. We do not aim to detect the
unvocalized laughter in this work.

% \note{1.Why the technique could not be applied for ``positive'' apex
%   detection?}

% \note{2.We can maybe merge II.A and II.B to ``The positive vocalized
%   laugh''. Should we use ``vocalized'' or ``apex''?}
  
%  \note{Sucheta on 260815: both are hard, need to discuss about that.}

\subsection{Rhythm Analysis}

Rhythm is defined as the systematic temporal and accentual patterning
of sound. In music, rhythm perception is usually studied by using
metrical tasks. Metrical structure also plays an organizational
function in the phonology of language, via speech prosody or laughter
\cite{clynes1982}. We attempt to use this metrical structure of the
human laugh without analyzing speech prosody. From the earlier
studies \cite{patel2006,sethares2007}, we see that prosody and musical
structure (such as rhythm) borrow or share concepts since long
back. This studies with rhythmic patterns lead to the birth of the
linguistic theories of stress-timed and syllable-timed languages. Here we
do not consider the rhythm in the speech prosody.
    
Recent studies \cite{bouzon2004,huss2011} reveal that ``rhythm'' in
speech should not be equated with isochrony. The absence of isochrony
is not the same as the absence of rhythm. In  \cite{bouzon2004},
isochrony is defined as the organization of sound into portions
perceived as being of equal or unequal duration. Strict isochrony
expects the different elements to be of exactly equal duration,
whereas weak one claims to have the tendency for the different
elements to have the same duration. So, the languages can have
rhythmic differences which have nothing to do with isochrony. But the
rhythm in human laugh is always isochronous like any music, so we
exploit the isochronous behavior of the human laugh in this work,
and do not consider the non-isochronous rhythm of languages.

We use an approach of frequency modulation to retrieve this rhythm,
following \cite{sethares2007,rn2014}. To detect rhythmic laughter
first we segment the whole speech to select the probable laughter
segments, then we classify the candidate frames for voiced laughter
using a rhythm algorithm based on frequency demodulation; finally we
select the rhythmic laughter frames through a statistical process. We
do not consider shared laughter captured on a single channel, rather
our method is engineered for a solo laughter by the single participant.
    
\section{Proposed Method}
\label{sec:method}

\subsection{Rhythm based laughter detection}

We use an unsupervised algorithm to detect laughter using its rhythmic
property. This entire process can be divided into three basic
sub-processes: first we filter out low power spectral density frames
using an automatic PSD threshold computation. Then we use all
high-energy PSD frames to detect all rhythmic segments (such as
rhythmic speech and/or rhythmic laughter) with the rhythm detector
algorithm. Finally, we compute a statistical threshold to detect only
the rhythmic laughter frames.

Based on the detected rhythmic laughter frames, we are able to
generate the time boundaries of the laugh segments. Description the
three aforementioned sub-processes is following.
 
 %{\bf (1) PSD Threshold computation: }
\subsubsection{PSD threshold computation}\label{psd} We compute the PSD threshold
using nonparametric power spectral density (PSD) estimation through
Welch's overlapped segment averaging PSD estimator $S(e^{j\omega})/F$,
where $F=fs$, i.e. the sampling frequency \cite{proakis1996}.
 
We compute the PSD threshold $D_{th}$ following the Otsu method
\cite{otsu1975}. In this method the computed PSD set (PS) is sorted in 
ascending order, let us consider the index sets as $[1\cdots L]$, then
the sorted set is divided into two sets randomly, say:  $\{1\cdots
k\}$ and $\{k+1\cdots L\}$, where $L=n(PS)$. Next, for $1<k<L$, we iteratively compute
$\sigma_{B}(L)$ then finally we compute $D_{th}$, given
by,  $$D_{th}=\max{\sigma_{B}(L)}=\max\big{[}\max(\frac{[\mu(L)*\omega(k)-\mu{k}]^2}{\omega(k)*(1-\omega(k))})\big{]}$$
where, $\omega_k=\sum_{i=1}^{k}prob_i$,
$\mu_k=\sum_{i=1}^{k}i*prob_i$, and
$\mu_L=\sum_{i=1}^{L}i*prob_i$. 
Here $prob_i$ denotes the $i$-th probability considering the elements of the corresponding set in the iteration, further details is in the paper by Otsu (1975)\cite{otsu1975}.

% \note{You have to describe all symbols you use: what is $L$,
%   and all those means and probabilities?--addressed on 260815}

%Alternatively, the PSD threshold computation can be done through an
%iterative process of running our laugh detection algorithm on the
%development data. In the section \ref{sec:res}, we experimentally
%compare the performance of PSD threshold computation from the
%development data in against to that proposed above through the
%unsupervised Otsu method \cite{otsu1975}.

We attempt to acquire the optimal value PSD threshold through a
brute-force optimization process of running our laugh detection
algorithm on the development data. In the section \ref{sec:res}, we
experimentally compare the performance of PSD threshold computation
with the development data using the unsupervised method by Otsu (1975)
\cite{otsu1975} and the brute-force optimization method.

  % by \cite{otsu1975}. 

% \note{I don't understand ``alternatively'', to what? We run the
%   threshold calculation always on the same data we detect. Or no?--addressed on 260815} 

 %We observe that the threshold is almost the same for both the cases. %%%%%%%%Sucheta Comments on 11/6: for some conversations this last statement is true but not for all, especially ICSI data is harder to analyze, as I already observed that when the speech signal is clipped our devised algorithm is having great trouble for detection.%%%%%%%%%%%

%{\bf(2) Rhythmic frame selection: } 
\subsubsection{Rhythmic frame selection} \label{sec:rframeselect}

First we select the high PSD frames using the threshold computed in subsection 
\ref{psd}. Then these high PSD frames are passed through the rhythm
calculation,
% (algorithm \ref{rhythm}), 
thus we select the rhythmic frames among all
the high energy frames. More specifically, we call these rhythmic
frames as the candidate laughter frames. %Therefore, in this method our features are the input signal $x$ itself and the power spectral density (PSD) estimate, $pxx$, of the input signal, $x$.

% \begin{algorithm}
% \caption{Rhythm algorithm}\label{rhythm}
% \begin{algorithmic}[1]
% \Procedure{rhythm}{$w_i$}
% \State{ToDo: convolve output vector $\hat{o}$} 
% \State{To initialize: bandlimits} 
% \State $\hat{y}=\text{fft}(w_i)$ 
% \State To find band ranges using band limits and $\hat{y}$
% \State To format $\hat{o}$ using band ranges and $\hat{y}$
% \State To compute $\hat{o}$: ifft(Hanning FFT * FFT($\hat{y}$)) 
% \EndProcedure
% \end{algorithmic}
% \end{algorithm}

% \textit{Rhythmic Pattern Detection: }

We basically exploit frequency modulation (FM) technique to capture
isochronous behavior of rhythm \cite{sethares2007}. In this case we
use an oscillator to modulate the frequency of a sinusoidal wave. Here the
oscillator is the ``carrier'' and the other one is the
``modulator''. We attempt to use a sawtooth carrier in this
case. Since laugh signal has a periodic nature it is traceable as a
sawtooth (or triangular) waveform, therefore we choose the triangular
hanning window as the basic oscillator function, which is computed as follows:  
 $$s=\Big[\cos^{2}(\frac{2*i*\pi}{l})\Big]_{i=1\cdots6}$$
here $l$ denotes the hanning window length. 

The properties of the ``modulator'' FM components are defined by the
frequency band limit with a set of six harmonics that starts with zero
then it reaches the periodicity pitch 200 Hz then all the other four
($2\times200$, $4\times200$, $8\times200$ and $16\times200$) harmonics
of that pitch. Here we choose to follow this filterbank implementation
method described in Scheirer(1998) \cite{scheirer1998}. Each harmonics
has two band-ranges. Therefore, this 
also initializes twelve band-range values. These frequency band-limits
are used to compute the band-ranges. Since the beginning of the method
we were computing data in the time domain. Now the signal is taken
from the time domain to the frequency domain with Fourier transform,
and we prepare the output 
using short time windows
%, half from back and half from the forth
. Finally, we convolve the inverse fast Fourier transformed
window data with a Fourier transformed half-hanning window.

% Subsequently in the
% procedure (3), we choose the frames of negative gradient (see details
% in procedure 3) as our candidate frames for laughter, like in the case
% of sawtooth carrier every even frame is used.
 
We use a set of six band-limits at this moment: beginning at 200 Hz,
increasing this in multiple of two, as the frequency results in a more
and more complex multi phonic. The resulting wave is the summation of 
many different sinusoidal waves; the carrier frequency lies in the
middle while the other tones lie above and below it at distances
determined by the modulation frequency. When the modulation amplitude
rises, the amplitudes of the additional frequencies also
rise. However, this increase is difficult to formulate
mathematically. The advantage of FM over additive one (the simple
addition of sinusoidal waves) is that we need to use only two
oscillators to convolute a rich and complex rhythmic human laugh
sound. Although currently we use the six modulation frequencies, this
number can be changed if needed. The output of this function is
basically a six column matrix, each row of the matrix is one frame. We use the median of this output to use
it further in the next subsection \ref{statp}.

% \note{rows of the matrix are the frames? -- Added one sentence beforeSG26.07}

% \note{I commented out the algorithm as it is well described by the text. "Ok -SG26.07"}

%{\bf(3) Rhythmic laughter frame(s) detection: }
\subsubsection{Rhythmic laughter frame(s) detection}\label{statp}

We compute the basic statistic functions (namely mean and standard deviation) for all the obtained rhythmic candidate
laughter frame with negative gradient (i.e. basically the negative
difference between two consecutive points, and this is done to compute
all local maxima points). Next, we derive the $95\%$-confidence bounds
through the Student $t$-test of the standard deviation. Then, we
compute a statistical threshold for rhythmic laughter frame selection
as the difference between the upper bound of the confidence interval
and the estimated population-standard deviation computed through same
Student $t$-test. We compute this threshold on the basis of the
hypothesis that the power of laugh is significantly higher than that
of rhythmic speech/music.
%there is a significant power based difference between rhythmic
%speech/music and rhythmic laughter. 
We select the frames as the
laughter frames whose standard deviation is equal or higher than the
threshold, and we finally compute the time intervals from those
selected frames.

% \note{Previous paragraph is very important, but I did not get
%   it. First, what are points and why do you calculate negative
%   gradient. Then, the statistical threshold for frame selection, what
%   is the link between the method and the the hypothesis you outlined?--addressed on the 260815, does this look clear now?}

%\begin{algorithm}
%  \caption{thres: PSD Threshold computation - process 1}\label{dth}
%  \begin{algorithmic}[1]
%    \State \textbf{Input:} Wavfile: $W={w_i}=\left(x_i,fs\right)$
%    \State \textbf{Initialize:} [wintime, steptime,T: noOfFrm, DTh=0]
%    \State \Comment{PSD threshold computation - process 1}
%    \State for{t = 1 to T}
%    	\State $x_i= {x_1,\ldots,x_n}$
%	\State px = welch\_periodogram($x_i$)
%	\State for {each set of px} 
%	  \State spx=sort(px); spx: $\{1\cdots L\}$
%	  \State iteratively to choose a value to divide spx: $\{1\cdots k\}$ $\{k+1\cdots L\}$
%	  \State compute $\omega_k=\sum_{i=1}^{k}prob_i$
%	  \State compute $\mu_k=\sum_{i=1}^{k}i*prob_i$
%	  \State compute $\mu_L=\sum_{i=1}^{L}i*prob_i$	  
%	%\EndFOR
%	\State $\sigma_{B}(L)=\max(\frac{[\mu(L)*\omega(k)-\mu{k}]^2}{\omega(k)*(1-\omega(k)))})$	
%    %\EndFor
%    \State Dth=$\max(\sigma_{B}(L)$);
%    \end{algorithmic}
%    \end{algorithm}

 Algorithm \ref{view} outlines overall view of the steps involved in
 the proposed laugh detection.

\begin{algorithm}
  \caption{Overall View of Laugh Detection}\label{view}
  \footnotesize{
  \begin{algorithmic}
    \State Input: Wavfile W
    \State read wavfile $\left(x_i,fs\right)=waveread(W)$
    \State $x_i=$data sample; $fs=$sampling rate
    \State Initialize: [frameSize, frameShift,T: noOfFrame, $D_{th}$=0]
   \State {\bf /*PSD threshold $D_{th}$ is calculated -- procedure 1*/}
    \State $D_{th}=$ threshold$(W)$   \Comment Otsu threshold computation  
   
       \State for$\{t = 1 \cdots T\}$ 
    	%\State $x_t= {x_1,\ldots,x_n}$
	 \Indent
	\State pxx = welch\_periodogram($x_t$) \Comment $x_t$: all data samples in $t$-th frame.
	  \State{ \bf /*rhythmic frame detection - procedure 2*/}
	   \State if ($pxx \ge D_{th}$) 
	    \Indent
	 	\State c=median(FMrhythm($x_t$)) \Comment  FMrhythm() detects rhythmic frame in six columns for six bands described in sec. \ref{sec:rframeselect}. In $c$ the median of the six bands stored.
		\State for (j = 2:length(c))
		\Indent
            		\State if ($c(j)-c(j-1) < 0$) $flg=1$;  \Comment local maxima check
		\EndIndent     		
        		\State endFor
        		\State $\sigma$=std$(x_t)$; \Comment std: the standard deviation
		\State if{(flg)} $\vec{\sigma}= [\sigma]$;  \Comment $\vec{\sigma}$ stores all candidate laughter frame stds $\sigma$
            	%\EndIf	
	 \EndIndent
	 \State endIf	 
	\EndIndent
	\State endFor
    \State {\bf /*laughter frame detection - procedure 3*/}
    
    \State if{(length$(\vec{\sigma})>0)$} 
    \Indent
	\State $(ci[lb,ub],sd)$ = ttest($\vec{\sigma}$)\\
	\Comment sd: the estimated population of standard deviation; $ci$: confidence interval with lower $lb$ and upper $ub$ bounds; uses 1-sampled t-test
	\State for({i =1:len($\vec{\sigma}$)})
	\Indent
        		\State if{$({\vec{\sigma}}(i)>=(ci[ub]-sd)$)} 
		\Indent
		\State Output: Print time interval of frame i.e. matching indices of $t$
		\EndIndent
	\EndIndent 
	\State endFor       		
       		%\EndIf
	\EndIndent
	\State endIf
    %\EndFor
    %\EndIf
  \end{algorithmic}}
\end{algorithm}

\section{Experiment}\label{sec:detexp} 
\subsection{Experimental Setup}\label{expset} 
In this framework, the method takes a raw speech signal as input and it outputs the
time intervals of the laughter segments in the signal. We consistently
apply a standard short-time analysis using a frame window of 2.5 sec
(following the study of \cite{ruch2001}) with 50\% overlaps. We used
part of AMI corpus as our test (5 meetings) and development (2 meetings) data.

\subsection{Data}\label{dati}
We used Augmented Multiparty Interaction (AMI) meeting corpus in this work.
   %\paragraph{ICSI meeting corpus}This corpus \cite{janin2003} is a collection of 75 meetings collected at the International Computer Science Institute in Berkeley. The meetings included are ``natural'' meetings in the sense that they would have occurred anyway: they are regular weekly meetings of various ICSI working teams. The collection includes in total of approximately 72 hours of Meeting Room speech. The audio was collected in multi-channel at a 48 kHZ sample-rate, down-sampled to 16 kHz on-the-fly.  In this work, we focused on the close-talking recordings. We did not work with the far-field recordings to detect simultaneous or shared laughter. We used first 23 Bmr folder data as the training split (baseline purpose), next 3 folders as development data (empirical study of the PSD threshold calculation) and the last three folders as test split, like \cite{kennedy2004,truong2007}. We used part of the data from the test split ( i.e. 120min recording in total; 2 meetings; 4 speakers: 2 male and 2 female).
%
AMI meeting corpus \cite{renals2007} consists
of 100 hours of meeting recordings. The recordings use a range of
signals synchronized to a common timeline. These include close-talking
and far-field microphones, individual and room-view video cameras, and
output from a slide projector and an electronic whiteboard. During the
meetings, the participants also have unsynchronized pens available to
them that record what is written. The meetings were recorded in
English using three different rooms with different acoustic
properties, and include mostly non-native speakers. Following
Petridis and Pantic (2011) \cite{petridis2011} we used only close-talk
headset audio (16kHz) 
recordings. We used the same data, which is used by Petridis and
Pantic (2011) \cite{petridis2011} (i.e. the seven meeting recordings
recordings of eight participants consisting 6 young males and 2 young
females of around 210 mins of recordings). We split the whole data set
into two parts: the development data consists of two meeting
recordings (i.e. IB4010 and IB4011 sets); we present the final result
shown in the Table \ref{bsln-rhythm-comp} using our test data of five
meeting recordings (i.e. IB4001 and IB4005 
sets). The challenge of AMI meeting corpus is that the data has a
large amount of overlapping speech. 
% \note{Sucheta on 260815: should we also give the specification
%   of the (main) experimental data or we remove the specification?}

\subsection{Baseline}
We follow the same baseline protocol using the same feature set like
\cite{salamin2013,kaushik2015}. We establish a baseline for the
(general laugh vs non-laugh)
classification of human laugh using Interspeech 2013 Paralinguistic
feature set. This feature set consists of 141 features
\cite{schuller2013}. We use support vector machine classifier with
5-fold cross-validation. We extract the features using the OpenSmile
tool \cite{eyben2010}. We use LibSVM \cite{chang2011} classifier for
SVM training and prediction. This is a supervised binary sequential
classification task. We use the data segments of 20 msec window at the
rate of 10 msec. The baseline is achieved in a speaker dependent
scenario. We select this baseline because it is the best performing
supervised method for laugh detection.

\subsection{Result \& Discussion}\label{sec:res}
% In this work we attempt to devise an algorithm for detection of
% positive vocalized laughter using rhythm, it is not devised for all
% kinds of vocalized laughter.
Table \ref{bsln-rhythm-comp} compares the results of our proposed
approach with the supervised baseline approach. While the proposed
rhythm based algorithm can be used for the detection of positive
vocalized laughter using rhythm, the baseline has been designed to
classify all kinds of laugh without distinction.
% \textit{This baseline also targeting to classify
%   all kinds of laugh without distinction}.
% We use AMI meeting corpus
% to evaluate our approach to the baselines using corresponding set of
% database.
We evaluate the results in the percentage F1-measures
\cite{powers2011}. The performance of our approach on AMI meeting
database shows better performance than the corresponding baseline. We
notice that it performs in a balanced way in terms of precision and
recall.
 % in case of the AMI meeting database. 

%The performance of our approach on ICSI database could not beat the performance of the corresponding baseline. The reason behind this is that the basic task of the baseline and the task of our approach differs: the baseline classifies the laughter versus the non-laughter, whereas in our approach we detect positive vocalized laughter against the others. The others imply to the other kind of laughters such as non vocalized satirist laughter and also the non-laughter segments of speech. There is another point to ponder on: the positive laughter seems to have a connection with the stress level. Among the two databases ICSI is the natural meeting corpus, and we know that at ICSI people work in a highly competitive environment. We speculate after error analysis that the impression of the stress at work is evident in the interactions in ICSI corpus with not many positive laughter, whereas the second best performer database AMI is recorded in a relaxed, role based meeting environment. \textit{The results of positive laughter detection may be faltered out with the increased level of stress in interactions captured across the two databases}.

% We depict the detailed performance of our approach on the AMI meeting
% database in the Table \ref{bsln-rhythm-comp}. We notice that it
% performs in a balanced way in terms of precision and recall in case of
% the AMI meeting database. 

\begin{table}
\begin{center}
\small{
\begin{tabular}{ l c r }
  \hline
  Data & Baseline & Our Approach \\ \hline
  %ISIL Data & 55.0 & 85.6 \\ 
  AMI Data & 81.1 & 84.5 \\  \hline \hline
%  ICSI Data & 97.6 & 74.3 \\ \hline \hline
\end{tabular}
}

\caption{Percentage F1-measure comparison between baseline \& proposed
  approach for test dataset split of our corpus} \label{bsln-rhythm-comp}
\end{center}
\end{table}

\begin{figure} \begin{center}

%\includepdf[pages={1}]{roc-crop.pdf}

\includegraphics[height=7cm]{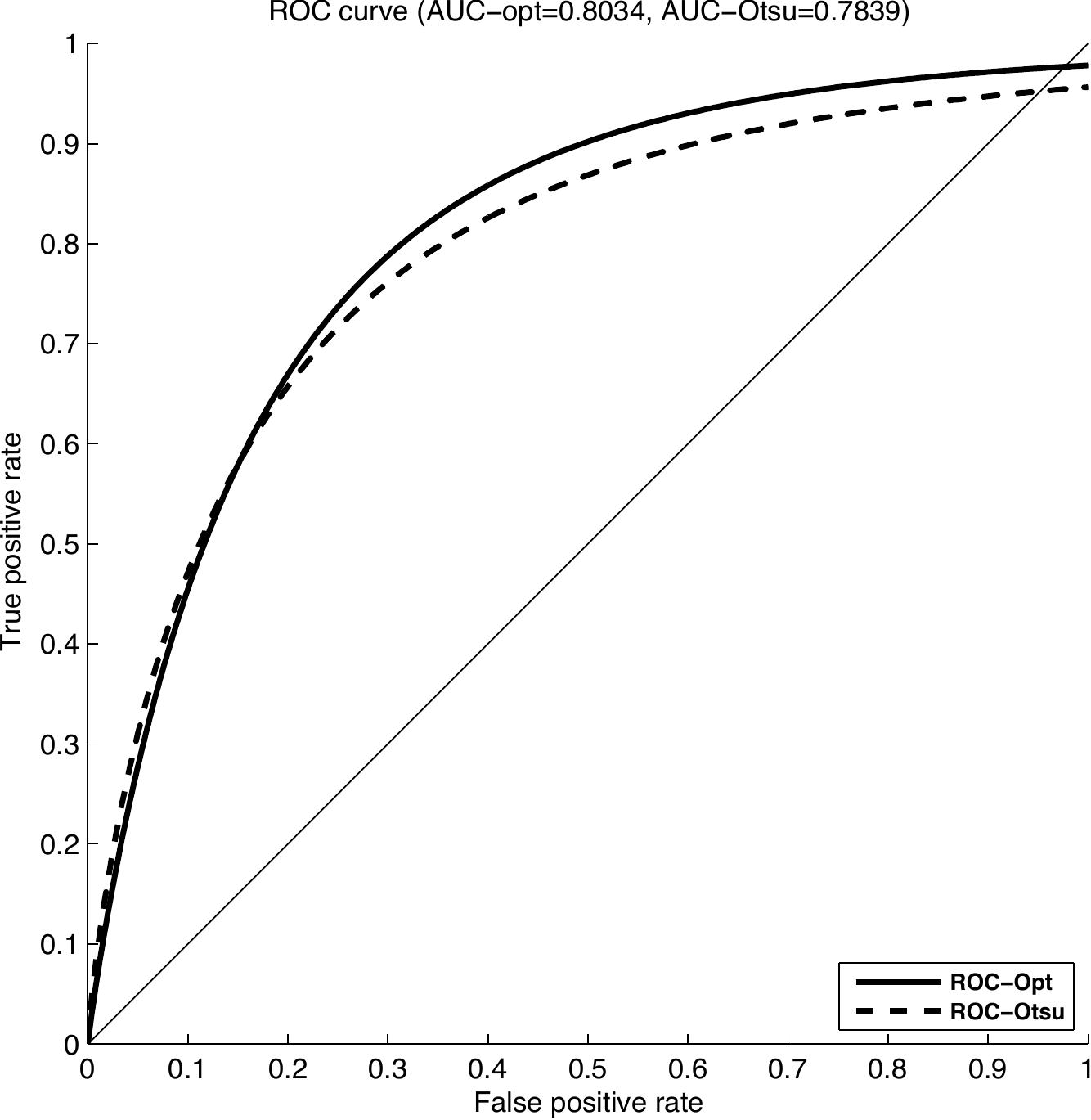}

\caption{PSD thresholding: ROC curve for Otsu thresholding and
  optimized thresholding with the development data} \label{resfig}
\end{center} \end{figure}

%%\begin{figure} \begin{center}
%%\begin{tikzpicture} 
%%\begin{axis}[
%%    ybar,
%%    enlargelimits=0.35,
%%    legend style={at={(0.5,-0.15)},
%%      anchor=north,legend columns=-1},
%%    ylabel={\#\%Scores},
%%    symbolic x coords={AMI,ICSI},
%%    xtick=data,
%%    nodes near coords,
%%    every node near coord/.append style={font=\tiny},
%%    nodes near coords align={vertical},
%%    ]
%%\addplot [pattern=north east lines] coordinates { (AMI,83.4) (ICSI,81.2)};
%%\addplot [pattern=dots] coordinates {(AMI,85.7) (ICSI,68.4)};
%%\addplot [pattern=grid] coordinates {(AMI,84.5) (ICSI,74.3)};
%%\legend{precision,recall,F1}
%%\end{axis} 
%%\end{tikzpicture} 
%%\caption{Detailed performance-histogram of our approach on the two databases} \label{resfig}
%%\end{center} \end{figure} 
Figure \ref{resfig} presents the ROC (Receiver Operating
Curve) comparison of the PSD threshold computation over the development
data. We compare the 
performance of threshold computation using the Otsu method \cite{otsu1975} in against
to that of the threshold computation using the brute-force
optimization method. The ROC is computed using the
true positive and false positive percentages. We see from the Figure
\ref{resfig} 
that the ROC computed with brute-force optimization thresholding is performing
marginally better than the ROC computed by the method of Otsu (1975)
\cite{otsu1975} with the development data. Therefore we use the
optimized threshold with the development data to present the result in the Table \ref{bsln-rhythm-comp}.

\section{Conclusions \& Future Works}\label{sec:concl}

In this work we have outlined a novel algorithm for positive vocalized
laugh detection using rhythm. This is a real-time, training-free 
approach in comparison to the existing supervised approaches of the
laugh detection. The algorithm is based on the rhythmic transforms
in laughter. The rhythm is analysed through frequency modulations
using a modulator and a sawtooth carrier. All the six carriers are
fixed, beginning at 200 Hz and the other four multiples of 200Hz. The
strength of this technique also resides in that: we do not need to
extract pitch or other intricate feature set to analyze rhythm; since
it does not involve any complex process of computation, or
intermediate file or memory handling, the time and space complexity of
this method is low. We used AMI-role based meeting dataset to evaluate
the proposed algorithm. The proposed laugh detection approach works
well in comparison to the supervised baseline.
  
In this work we do not detect all kinds of vocalized laugh, we focus
on the detection of rhythmic vocalized human laughter. This method is
capable to work incrementally for further recognition of different
laugh types or other recognition for laugh and rhythmic speech or
music.%This method may be
%effective for spoken dialog tasks. It is capable of producing better
%results if the data is non-overlapping speech. %This approach may also
%be applicable to the e-learning scenarios, to account the students
%attention towards the ongoing lessons. Again, it is possible to use
%this laugh detection technique as a procedure of implicit annotation
%or highlight detection in speech, to reduce the cost and time
%complexity for human annotation. 
This algorithm is sensitive to speech
signal clipping; it may fail to detect the human laugh recorded in a
noisy environment, specifically, it will fail to work with human
speech data along with a background score of rhythmic music.

The Matlab code of the algorithm is available as open-source code at
the following address: \url{https://github.com/sghoshidiap/LaughDet}.

 \section{Acknowledgements}
  
\scriptsize {
We thank all the people who helped this work at ISI Calcutta India,
specially Dipabali Sarkar, Madhupa Das Bairagya and Sairik
Sengupta. Special thanks go to Afsaneh Asaei of Idiap Research
Institute, Switzerland and Yves Leprie, Loria, France for critical reviews.}
\newpage
\balance
%\vfill\pagebreak

%\section{REFERENCES}
%\label{sec:refs}

%List and number all bibliographical references at the end of the
%paper. The references can be numbered in alphabetic order or in
%order of appearance in the document. When referring to them in
%the text, type the corresponding reference number in square
%brackets as shown at the end of this sentence \cite{C2}. An
%additional final page (the fifth page, in most cases) is
%allowed, but must contain only references to the prior
%literature.

% References should be produced using the bibtex program from suitable
% BiBTeX files (here: strings, refs, manuals). The IEEEbib.bst bibliography
% style file from IEEE produces unsorted bibliography list.
% -------------------------------------------------------------------------
\small{
\bibliographystyle{IEEEbib}
\bibliography{IEEEexample}}

\end{document}